30th CIRP Design 2020 (CIRP Design 2020)

# Towards automated Capability Assessment leveraging Deep Learning

Raoul G. C. Schönhof*[a], Manuel Fechter[a]

[a]*Fraunhofer Institute for Manufacturing Engineering and Automation IPA, Nobelstraße 12, Stuttgart, Germany*

* Corresponding author. Tel.: +49-711-970-1843; *E-mail address:* raoul.schoenhof@ipa.fraunhofer.de

**Abstract**

Aiming for a higher economic efficiency in manufacturing, an increased degree of automation is a key enabler. However, assessing the technical feasibility of an automated assembly solution for a dedicated process is difficult and often determined by the geometry of the given product parts. Among others, decisive criterions of the automation feasibility are the ability to separate and isolate single parts or the capability of component self-alignment in final position. To assess the feasibility, a questionnaire based evaluation scheme has been developed and applied by Fraunhofer researchers. However, the results strongly depend on the implicit knowledge and experience of the single engineer performing the assessment. This paper presents NeuroCAD, a software tool that automates the assessment using voxelization techniques. The approach enables the assessment of abstract and production relevant geometries features through deep-learning based on CAD files.
© 2020 The Authors. Published by Elsevier B.V.
Peer-review under responsibility of the scientific committee of the CIRP Design Conference.

## 1. Introduction

For most manufacturing companies, assembly represents the core area of value creation and cost [1]. It faces daily challenges to master the balance between quality, required quantity and the limited availability of resources in production. The automation of assembly processes is one solution to create high productivity, especially in high-wage countries with limited or expensive human resources. Permanently, the decision to automate or to retain the manual assembly process is challenged by the technical suitability and feasibility as well as economical considerations. The question of whether components or assembly groups can be assembled automatically, as well as the question for the appropriate level of automation [2] are often asked during industrial consulting projects. For this purpose, Spingler and Beumelburg [3][4] developed an automation score system based on a questionnaire, assessing relevant aspects of the component separation, material handling, positioning and joining processes. Based on the performed assessment, a relative score is obtained, suggesting whether the assembly solution should be performed manually or automatically.

Although the questionnaire is easy to answer for an experienced automation engineer, it is currently too complex to be automated within a simple rule-based computer program. In the context of this work, multi-layered neural networks (so-called deep learning networks) are applied, which suggest problem solving capabilities for the described use case. Although neural networks were first proposed in the early 1940s of the 20th century and can be regarded as programmatically controlled since the 1980s [5], it took another 30 years before they were widely used in science and industry. Mainly due to the increasingly efficient computer systems, neural networks can reach a level of complexity that even enables them to partially comprehend human mental tasks. The described technology plays to its strengths especially in recurring activities. This leads to the assumption, that Deep

Learning could likely be used effectively in the context of the described automation assessment.

Section 2 of this paper provides an overview of the state of the art and work already done in the field of automation assessment. Section 3 introduces the *NeuroCAD* software, a web-based service that provides users with various information about the technical feasibility of automation considering a product part component in just a few seconds. Section 4 discusses the evaluation results, while Section 5 summarizes the paper. Section 6 provides information on upcoming work.

## 2. State of the Art

The outlined state of the art distinguishes between two main disciplines. On the one hand side, the assembly process itself and the procedure of identifying automation potential as seen from a manufacturing engineer's perspective. On the other hand, the section illustrates how (artificial) neural networks are built, how they work and how this kind of artificial intelligence can be used to solve complex tasks that rely on implicit process knowledge like the mentioned automation assessment. The section ends with an overview about existing work in the field of computer-based automation assessment.

### 2.1. Aspects of Assembly

From an operational point of view, assembly occupies a special position during production as most of the added value is performed during these steps. Most products reach their final functionality and quality once they have been assembled and aligned. The automation capability describes the technical feasibility of a process to be fully or partially automated by machines. The question whether an assembly task can be automated or not depends on e.g. the geometry of the component, the type of assembly process and the material itself. Therefore, the component characteristics to be investigated during the mentioned assessment ranges from geometric over process-related up to physical properties of a component and their functional relations to adjunct parts in the assembly group. Table 1 highlights some key aspects of the automation decision.

Table 1: Exemplary characteristics to assess the automation score

| Focus Area | Examples |
|---|---|
| single part | material rigidity; existence of gripping surfaces and orientation markers; availability of joining aids; surface sensitivity; material supply; length ratios |
| assembly groups | applied joining technology; tolerances; accuracy; repeatability; accessibility to the joining positions |
| assembly process | self-locking mechanism in final position; availability of the automation technology; concatenation of the process steps |

The assembly process follows the characterization into two major domains of handling and joining [6][7]. The domain of handling is separated into the sub-domains of material transport and manipulation, the division and separation of parts from a larger quantity and the positioning of parts relative to each other. The joining domain is separated into different joining technologies depending on the physical principle and the type of rigid or detachable connection.

The assembly process usually starts with the task of separation. The difficulty of this task is primarily determined by the material supply and the shape of the part. The decision whether a component can be easily separated from bulk material or requires special machinery relies on implicit background knowledge about the availability of separation devices and their used separation technology. A component with simple geometries and existing symmetries can be easily separated using mechanical principles e.g. shaking or spiral conveyor systems. In the case of sheets and block parts with more complex geometries, it may be advisable to separate them manually or with sophisticated bin pick technology. This implies that suitable gripping surfaces are available at the component body. It is also possible that there is no automatic separation solution available on market.

Handling and manipulation is a crucial function of the material flow. It is intended to manipulate and position the objects during different assembly steps. In general, the material stiffness, the existence of gripping surfaces and orientation features, as well as surface sensitivities are relevant for the evaluation of the handling attributes.

The assessment of the positioning characteristics includes the required accuracy of the target position, existing positioning aids, like chamfers or fillets on the target parts, the local accessibility to the joining position, the joining motion, the corresponding joining tolerances and the ability to self-lock and stabilize in the final position. In comparison to the previously discussed assembly aspects, the positioning criteria cannot be achieved by looking isolated at one individual component. An overall consideration of adjunct parts is necessary, which includes the relevant components for this particular assembly step.

Depending on the joining process, additional space may also be required to access the joining position with required tools. The joining property is mutually determined by the component geometry and the assembly technology and process. The evaluation of the joining process for two individual components can hardly be determined based on fixed rules. It therefore requires higher implicit knowledge.

### 2.2. Computer-based Automation Assessment

Current automation assessment is performed manually [8]. Automated procedures are limited to the recognition of geometrical characteristics of *computer-aided design* (CAD) components, so-called feature extraction. Most approaches do have in common that an information representation of the parametric CAD model is performed by compression. Ozturk suggests defining the input vector over the corners, edges and surfaces of the component so that the input vector contains a pre-compressed representation of the model [9]. These efforts led to the adaptive resonance theory (ART), which attempts to transform geometric properties of a CAD model into a quantitative description of the surface elements [10]. Qin and Su, on the other hand, tried to avoid such an abstract information representation by using a two-dimensional projection. The 2D images derived from the 3D CAD models could thus be processed more easily via a (standard) convolutional neural network (CNN) [11][12]. In both approaches, the aim of the systems was to recognize and categorize similar components.

The approach to process input data via the CAD parameters is also used in current systems for model decomposition, i.e. the decomposition of a complex model into primitive forms [13] respectively clustering [14]. A noticeable common feature of all approaches is that the model data has to be simplified or strongly abstracted. In particular, the wish to use a 2D projection of a 3D model is excellent to classify a model in terms of construction type. However, it should be difficult to learn spatial manipulations with this procedure.

Although it is not in the field of automation assessment, another work is relevant to this topic. It takes the principle ideas depicted at Chon et al. as a part of the evaluation of magnetic resonance imaging data. Chon et al. presented a system that allows different types of lung cancer to be classified by examining 3D data, using a CNN auto encoder [15]. Even if the data is represented as points in a 3D space, it is not a conscious decision by Chon et. al., but merely a result of the magnetic resonance imaging data and its image processing. On the one hand, information compression can be achieved by reducing the resolution without destroying the 3D structure of the model. On the other hand, this structure allows the application of 3D convolutions.

## 3. Proposed Work: NeuroCAD

The paper presents the *NeuroCAD* [16] system that supports the automation assessment of an existing product part component. The system performs the assessment autonomously for a given assembly feature based on input CAD data by recognizing and interpreting shapes. It is advantageous solely to rely on CAD data due to the missing interaction with an expert. At the same time, it is possible to link the system to the product data management systems (PDM systems). This offers a direct access to the CAD data and corresponding classifications. In particular, the interface to the PDM systems allow a direct assessment of a large number of models, which facilitates the identification of automation potentials on multiple parts.

In addition, the PDM integration allows the design department to easily access production planning expertise by iteratively checking the feasibility of automated assembly during the design process. The result is a system that automatically answers many of the questions about assembly automation, described in Section 2. The system response should be indirectly traced back to the component geometry. Two situations have to be distinguished in the context of assembly. On the one hand, a component can be unfeasible for automated assembly, e.g. if it cannot be automatically separated from material supply. On the other hand, two components cannot or can only be joined unfavourably, e.g. if the joining has to be carried out by a complex path.

Within the framework of model-specific analysis, a single model is assessed, investigating properties that do not show interdependencies with other components. These include the ability to automatically separate the component from material supply, the presence of gripping surfaces and orientation features for accurate gripping and handling. The analysis of assembly groups focuses on the tasks of positioning and joining, therefore on properties, which are characterized by mutual dependencies between two components.

### 3.1. Machine Learning System Architecture

The core of the automatic assessment service was a CNN, which receives the CAD data as input variables and projects them into a defined response. One component of the CNN was the "training data generator", which provided the corresponding input and output variables from a manually annotated dataset. The input data contained the processed CAD model in a form suitable for the input neurons. The output data contained the answers, i.e. an expert's assessment to be learned by the neural network, with regard to the respective component properties.

Creating the dataset, CAD models from Fraunhofer database as well as from public CAD download platforms, like "Trace Parts" [17] were used. When dealing with CAD models as input vectors, the information of the models has to be considered. The models are available in a parametric format, such as *Standard for the Exchange of Product model data* (STEP), or tessellated, e.g., *Standard Tesselation Language* (STL) or *wavefront* (OBJ). The designer has the advantage that the models can be scaled arbitrarily without losing the information about the model and its internal relations. However, such models are not directly suitable for being fed into a neuronal network, since the component geometry is not fixed like it is with e.g. photo image points. The projection of a parameterized model rather has to be calculated for each angle of view.

Finally, a neural network can only process data in the shape of a numerical vector. To convert the parameterized raw data of the CAD model into an appropriate form, the input-side interface has to be defined. A 3D tensor was considered suitable as input model representation for NeuroCAD. The input model thus represented a body that spans the three dimensions of spaces.

To transfer the raw data into the required 3D tensor, a conversion algorithm was used. Existing scripts are applied for this conversion. Although no tools exist that allow a direct STEP to 3D array conversion, it was possible to convert a STEP file into the STL format [18] by using the "aocxchange"-library. The so gained STL file was converted into a voxel model using the Binvox Library [19][20]. After the voxelization, the model only described a boolean 3D model. In such a model, each cell in the array contained either *True* or *False*. While the *True* value describes that the voxel grid contains the CAD model at the location, a *False* value describes that there is no model. Thus, the latter value is to be interpreted as "empty space". To arrive at the desired numerical array for a neural network, each Boolean value (*True*/*False*) was finally substituted by a float value (1.0/0.0). An example of a rasterized CAD model is given in Figure 1.

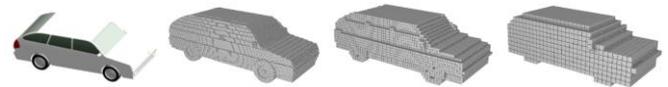

Figure 1: CAD model of a car at different voxel resolutions. From left to right: Original from Trace Parts [17], rasterized model at $180^3$, $64^3$ and $32^3$ voxels per model. Visualized by Drububu Voxelizer [21]

In order to train the network in a supervised way, labelled output data is required. For this purpose, an annotation interface was implemented to ask automation experts about the properties of a given model to be investigated (annotator with GUI). The software enabled the annotation for single components and assemblies.

*NeuroCAD* attaches particular importance to the modelling and representation of the output. In particular, it should be noted that during training, the network was likely to imitate both the target output and the error, or the quality of the imitation, based on the target and actual output data. In the following, the choice of data representation is explained using the question of separation of components as an example.

If an expert was asked how good a single component can be separated from a quantity of components, he could currently choose from the gradations *easy*, e.g., separable with a bowl feeder, *difficult*, e.g., separable by bin picking or *not separable at all*. The problem with only three gradations is that the probability of a randomly correct hit of the neural network would be one third. Therefore, a finer gradation in eleven steps was introduced into *NeuroCAD*, decreasing the probability of random correct answers to 9.09 %. Facilitating the training process, the specific activation energy E of each output neuron I, depending on the answer A, is defined by the relation of:

$$E_I = e^{-(0.5*(\frac{A}{11}-I+1))^2} \quad (1)$$

Figure 2 shows the activation of the neurons 0 to 10 for a score of 5, representing a medium difficulty for separation.

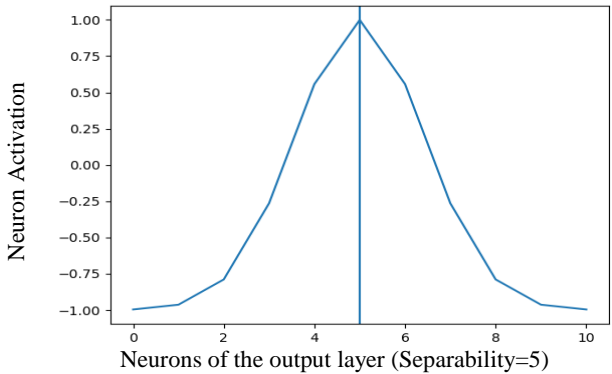

Figure 2: Intended output with a medium degree of separability

After all input data has been annotated with the desired output data, it was used to train the network. However, it must be taken into account that the provision and annotation of tens of thousands of models requires a lot of time. One way to generate a large training set from a small number of initial models is to create invariants of the initial models. The voxel model is rotated and scaled around all dimensional axes while maintaining its proportions. The models obtained by this approach represented a completely new input array from the point of view of the neural network, since filters already learned cannot be applied one-to-one to the invariant. Thus, a multitude of new models were automatically created from a single input model. Even if these models do not contain any new shapes, they still enabled the mesh to learn three-dimensional structures more easily by forcing it to reprocess the twisted model.

*NeuroCAD* was implemented using Tensorflow, which simplifies elementary functions such as creating tensors and performing arithmetic operations. In addition, Tensorflow provides functions that include optimization algorithms such as ADAM [22] and applies directly to weights and biases.

*3.2. Architecture of Neural Networks*

An overview of the network is shown in Table 2. The network consisted of eleven layers, of which six layers for *Feature Extraction* and two layers for *classification*. In the following paragraph, the net type dealing with the analysis of individual parts, especially on the attributes of separation and handling, will be explained in detail. The *Feature Extraction* component consisted of a CNN with eight layers, whose input layer held a model with a grid of 64x64x64 voxel and thus consisted of 262.144 neurons. The output after the eighth convolution layer consisted of 512 neurons, therefore 512 different abstract properties could be distinguished. From the first convolution to the last, the data undergoes different transformations over eight layers of different thickness.

For those networks, analysing the assemblies and therefore especially the attributes of positioning and joining, a different approach was used. Three CAD component models, consisting of two individual parts and the final assembled configuration, were combined to one input model with the dimensions of 192x64x64. In consideration of a relatively small set of training data, the number of layers was reduced to six in order to counteract a possible overfitting during optimization. In contrast to the *Feature Extraction* component, the classifier was always designed identically. It consisted of two fully networked layers, which used the sixth convolution as an input layer. The results were transferred to a layer with 128 neurons and to the final output layer of eleven neurons.

Table 2: Layout of the CNN network

| Layer | Type | Size out | Filters | Filter size |
| --- | --- | --- | --- | --- |
| Input | - | 64x64x64 | - | - |
| 1 | CNN+Pooling | 32x32x32 | 32 | 2x2x2 |
| 2 | CNN+Pooling | 16x16x16 | 64 | 2x2x2 |
| 3 | CNN+Pooling | 8x8x8 | 128 | 2x2x2 |
| 4 | CNN+Pooling | 4x4x4 | 256 | 2x2x2 |
| 5 | CNN+Pooling | 2x2x2 | 512 | 2x2x2 |
| 6 | CNN+Pooling | 2x2x2 | 512 | 2x2x2 |
| 7 | CNN+Pooling | 2x2x2 | 512 | 2x2x2 |
| 8 | CNN+Pooling | 1x1x1 | 512 | 2x2x2 |
| 9 | Fully Connected | 128 | - | - |
| 10 | Fully Connected | 11 | - | - |
| Output | - | 11 | - | - |

*3.3. Training Parameters*

The implemented cost function described the area between the target and the predicted curvature. It has to be admitted that training data was usually not perfect. There is always the possibility of incorrect annotated data records. Especially in the case of expert annotation with different implicit knowledge, a deviation can occur. A deviation of up to +-20% was estimated as a permissible deviation, e.g., a score of 5 instead of 3, depending on the human individuality of assessing situations. A higher deviation would be considered as an error during training. In analogy to the Boolean cases false-positive and false-negative, a deviation of more than 20% could be caused either by an error of the network or by an error of the expert input during labelling. The problem was, that the training algorithm could not detect an error during optimization. In contrast, the training algorithm would try to imitate the errors of the expert. To encounter the efforts of the training algorithm, sanctions by the cost function have to be way bigger than small errors. This goal was achieved by the implemented cost function C, which itself iterated elementwise over $i$ on the expected output $x_{exp}$ as well as the predicted output $x_{pred}$.

$$C = \sum_{i=0}^{10}(e^{-x_{i,exp}^2} - e^{-x_{i,pred}^2})^4 \quad (2)$$

The weights were manipulated by the *ADAM* optimization algorithm [22]. It has been shown for the training tasks, that a permissible change rate between 0.01 - 0.05% achieved high training success. At the same time, the training duration was limited to a bearable period of one to two days with two applied *Nvidia GTX 1080 Ti*. If the value was further reduced, no relevant improvement had been achieved. At the same time, the duration of the training was drastically increasing, as the cost function converged slowly towards zero.

**4. Evaluation**

Within this project, 207 models were annotated by experts. Out of these, 20 randomly picked models were selected for the evaluation set. The remaining models were used to create a total of 22,440 invariants by the described data augmentation. In case of accuracy, the percentage of actual output data which corresponds to the target output was continuously determined. An example for a correct respectively faulty system response is depicted in Figure 3 and Figure 4.

The untrained net had an accuracy of 5%. This corresponds to the expectation of an untrained network, which would randomly hit a correct value with a certainty of 1 out of 11. After 60,000 eras, the net assigned 100% of the training data identically for the first time. After 160,000 eras, there was no further improvement in accuracy. However, when a new batch was loaded after epoch 65,500, the accuracy dropped to 24% only. Due to the black box character of the network, the reason for this behaviour can only be guessed.

A possible explanation is either an extremely unfavourable variation of the weights by ADAM or a temporary overfitting during training for the current training data set. The specialization gained allowed the network to achieve extremely low costs for the actual batch, but after the change, it proved to be worthless. It could be shown that the network was able to handle 95% of the previously unknown models within the permissible interpretation range of 18% correctly. The maximum error was 27% and occurred at one model.

The results of the analysis on gripping surfaces showed a comparable outcome. In relation to the feasibility of separation, the detection and identification of gripping surfaces proved to be a relatively easy property to learn. In contrast to the study of separation attributes, it was possible to correctly evaluate all evaluation models with a tolerance of +-18%. The accuracy aimed for a proper value of 90%. The maximum error between the evaluation of the expert and the evaluation of the network was a tolerable 18% for all models of the evaluation set.

While the training for the single question of "self-centering behaviour during the closing of handling devices" showed no obvious results, the average accuracy was 70% within the permissible interpretation range of 18%. The maximum accuracy during the evaluation in particular seems to be disproportionately good. At times, the evaluation of the self-centering criterion turned out to be much better than expected (100% accuracy with a tolerance of 18% and 90% accuracy with a tolerance of 9%) for a rather complicated question with only a total of 200 training models. That is the reason why the result should be critically questioned.

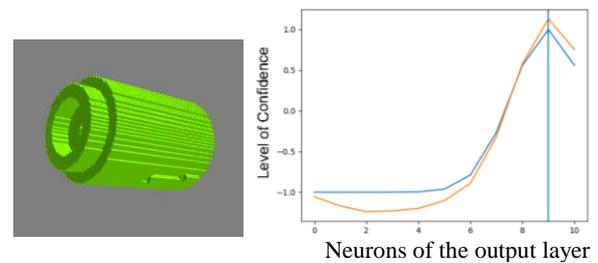

Figure 3: A correct prediction. The input model (left) provoked a prediction (right, orange curve), which corresponds to the expected response (right, blue curve)

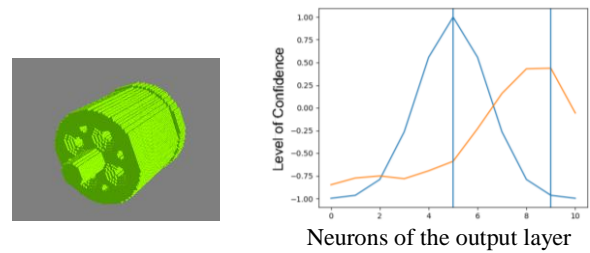

Figure 4: A false prediction. The input model (left) provoked a prediction (right, orange curve), which is off in relation to the expected response (right, blue curve)

*4.1. Evaluation of the Level of Confidence*

The idea behind the concept "*Level of Confidence*" was the minimization of the area within the predicted and expected curve. This working principle matches the assumption that an unknown and poorly interpretable input model leads to a curve, which differs strongly from the intended ideal bell curve shape and height. The less the model fits into the learned patterns, the more the maximum occurring value of the calculated curve will diverge off the ideal target value of 1.

This system behaviour was considered as a learned quality coefficient that not only contains the result itself, but also allowed a statement about the quality of the assessment. The network was trained seven times from scratch, each time with a different set of test data respectively evaluation data. The evaluation set contained 20 models, which lead to an accumulated evaluation set of 140 models. After the training, the derivation between the predicted and expected score was calculated. Additionally, the divergence between the expected height of the highest curve point (by default 1) and the predicted height was calculated. The relation is shown in Figure 5.

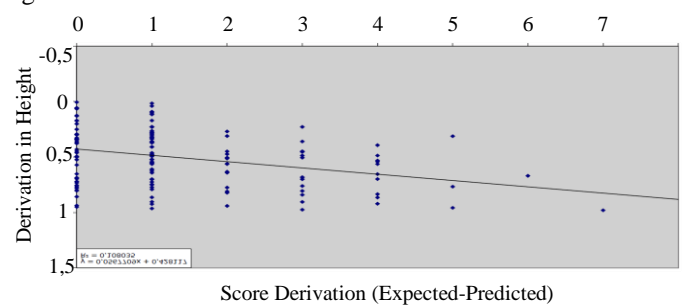

Figure 5: Relation between prediction errors and curve heights

The relation was approximated by linear regression, illustrated by the black line in Figure 5. The quality of the regression amount is determined to be $R^2 = 10.6\%$. The reason could be different process knowledge of the experts. Although it seemed promising, an evident correlation between the calculated and intended data-sets requires further investigations. Nevertheless, it showed that a correlation is feasible.

## 5. Conclusion

The approach showed that sub-tasks of the Fraunhofer automation assessment, can be automated using neural networks – a former complete manual task with high requirements into implicit automation knowledge. For this purpose, a system was implemented which analyses the geometric characteristics of the CAD models and answers questions of the automation capability questionnaire, such as the topics of separation, manipulation and handling, positioning and joining. The utilised convolutional neural networks achieved an accuracy of 80%.

A service system was designed and implemented, which allows to import common CAD models, to assess manual annotations and labels on these input data-sets in a time-efficient manner in order to create a training and evaluation data-set. The system was extensively evaluated in numerous learning sessions with more than 200 models. It could be shown that the performance of the system increased within the number of training models applied. Different types and parameters of training were tested and benchmarked. It could be demonstrated that the system recognizes geometric properties of a CAD component and interprets them correctly.

Indications were found that the system was not only able to evaluate models, but also returns a quality score, the so-called "Level of Confidence". This allowed the conclusion of the network certainty on its proper result.

## 6. Future Work

The system shows preliminary good results in the topic of automated automation capability assessment of CAD models to optimize manufacturing processes, e.g. the area of automated assembly. It offers potential to continue research for optimizing the current approach. Therefore, it would be interesting to investigate how the system behaves with highly complex shapes or more complex assembly groups. To do so, the training set has to be increased by at least a factor of 10. It would also be of great interest to see how an increase in the voxel model resolution ($> 1024^3$ voxel) affects the system's accuracy. High performance computing methods could be used to investigate different model properties in a distributed system. Additionally, the system will have to become more transparent. The system has to visualize, which regions of the CAD part are associated with the particular decision. In the long term run, this information might also be used to optimize the part itself. Finally, semi-supervised learning approaches [23] as well as AR support for assessing real parts [24] should be considered.